\documentclass[referee]{raa}            

\usepackage{graphicx,times}             
\usepackage{natbib}
\usepackage{amssymb,amsmath}
\bibpunct{(}{)}{;}{a}{}{,}

\usepackage[pagebackref=true,colorlinks = true, linkcolor = green, anchorcolor = red, citecolor = blue, filecolor = red, urlcolor = red]{hyperref}

\begin{document}

   \title{Application of the space-based optical interferometer towards measuring cosmological distances of quasars.
}

 \volnopage{ {\bf 20XX} Vol.\ {\bf X} No. {\bf XX}, 000--000}
   \setcounter{page}{1}

   \author{Ying-Ke Huang\inst{1}, Yue-Dong Fang\inst{1}, Kai-Xing Lu\inst{2}, Zhi-Xiang Zhang\inst{3}, Ji-Lin Liu\inst{1}, Sha-sha Li\inst{2} Bao-Rui Luo\inst{1}, Qin Lin\inst{1}, Zhuo-Xi Huo\inst{1}
   }

   \institute{ \inst{1}	Qian Xuesen Laboratory of Space Technology, China Academy of Space Technology, Beijing 100081, People’s Republic of China; {\it huangyingke@qxslab.cn}, {\it huozhuoxi@qxslab.cn}\\
\inst{2}Yunnan Observatories, Chinese Academy of Sciences, Kunming 650011, People’s Republic of China \\
\inst{3}Department of Astronomy, Xiamen University, Xiamen, Fujian 361005, People’s Republic of China 
 }

\abstract{ Measuring the quasar distance through joint analysis of spectroastrometry (SA) and reverberation mapping (RM) observations is a new method for driving the development of cosmology. In this paper, we carry out detailed simulation and analysis to study the effect of four basic observational parameters (baseline length, exposure time, equivalent diameter and spectral resolution) on the data quality of differential phase curves (DPCs), furthermore on the accuracy of distance measurement. In our simulation, we adopt an axis symmetrical disc model of broad line region (BLR) to generate differential phase signals. We find that the differential phases and their Poisson errors could be amplified by extending the baseline,
while the influence of OPD (optical path difference) errors can be reduced during fitting the BLR model. Longer exposure time or larger equivalent diameter helps reduce the absolute Poisson error. Therefore, the relative error of DPCs could be reduce by increasing any of the above three parameters, then the accuracy of distance measurement could be improved. In contrast, the uncertainty of $D_{\rm{A}}$ ( absolute angular distances) could be improved with higher spectral resolution, although the relative error of DPCs would be amplified. We show how the uncertainty of distance measurement varies with the relative error of DPCs. 
For our specific set of model parameters, 
without considering more complicated structures and kinematics of BLRs in our simulation, it is found that the relative error of DPCs $<$ 20$\%$ is a limit for accurate distance measurement.
The relative error of DPCs have a lower limit (roughly 5$\%$) and the uncertainty of distance measurement can be better than 2$\%$. 
}

   \authorrunning{Y.-K.Huang et al. }            
   \titlerunning{Application of the space-based optical interferometer}  
   \maketitle

%

\section{Introduction}
\label{introduction}

One of the basic open questions of modern astrophysics is to accurately measure the cosmological distances of extragalactic objects to understand the increasing $H_{0}$ tension (\citealt{Freedman2010, Peacock1999, Freedman2017, Weinberg2013, Riess2019}).  
Active Galactic Nucleis (AGNs), which are known as the brightest objects in the universe, 
are utilized to measure cosmological distances since they were discovered (\citealt{Sandage1965, Hoyle1966, Longair1967, Baldwin1977, Elvis2002, Quercellini2009, Wang2013, Marziani2014}). 
Nevertheless, due to lack of understanding of AGN physics, many attempts have been made but none of them were proven to be bias free.
Recently, a joint analysis (\citealt{Wang2020}) of spectroastrometry (SA: \citealt{Bailey1998, Petrov1989, Gravity2017}) and reverberation mapping (RM: \citealt{Blandford1982, Peterson1993, Kaspi2000, Bentz2013, Du2018}) observations provide a  direct method to measure absolute angular distances ($D_{\rm A }$) of AGN. This method do not need calibration using cosmic ladders so it has the potential to provide absolute distance measurements for quasars in the high-$z$ Universe.

The high-energy output of AGN is argued to be the result of accreting matter by a supermassive black hole at the center of galaxy (\citealt{Lynden1971}). 
A hallmark of the AGN spectra is the existence of broad emission lines (\citealt{Osterbrock1986, Osterbrock1989, Ho2008}).
These broad emission lines are thought to originate from the photoionization of Broad Line Region (BLR) clouds by the high-energy continuum photons from the central accretion disc. 
These clouds revolve dominated by Kepler motion around the central black hole due to gravity, 
which has been confirmed by multiple campaigns of several AGNs (\citealt{Peterson2004}).
The emission lines are broadened up to several thousand $\rm{km\ s^{-1}}$(\citealt{Antonucci1993, Urry1995}). 
However, it is impractical to spatially resolved the BLR by direct photometry, because the angular size of the BLR is smaller than the existing spatial resolution limit
($<$ 0.1 mas; \citealt{Blandford1982, Elvis2002}).

In the past few decades, a promising approach to directly constrain BLR geometry is the widely used RM technique. 
The underlying principle of RM technique is the photoionization theory (\citealt{Wandel1999, Kaspi2000, Bentz2006}). 
By analyzing the variation of continuum and emission lines afterwords measuring the time delay between the variable continuum and the responsive broad emission line, it is possible to constrain the geometry and kinematic of BLR and further estimate the characteristic size of BLR as well as the black hole mass.
In 2018, \cite{Gravity2018} proposed another method.
By measuring the photocenters of emission lines in different wavelengths channels, 
SA method (\citealt{Beckers1982, Petrov1989, Petrov1992, Gnerucci2010}) shows great advantage in  providing information on the spatial structure 
at scale much smaller than the spatial resolution of interferometer. 
An intuitive picture of how SA method can 
spatially resolve the bulk motion of the BLR clouds 
is illustrated as the follow: 
imaging there are two clouds located at a distance significantly smaller than the spatial resolution, so they are seen as a non-resolved source with a global angular size, but one cloud is moving towards observers and the other moving away from observers.
As the line-of-sight (LOS) velocities of the two clouds are different,
the photons from the same emission line but from different clouds are shifted to different wavelengths.  
By measuring the interferometric phases (providing position information on-sky) over wavelengths 
,we can obtain the photocenter position of clouds with different velocities and further separate the motion of BLR clouds. 
SA method shows the capability in 
overcoming the spatial resolution limit,
thus being able to constrain the angular scale, geometric structure and dynamics of BLR.

In fact, RM observations provide the linear size of BLR 
and are more sensitive to the direction along the LOS. 
SA measures the angular size of BLR 
and is more sensitive to a direction perpendicular to the LOS (\citealt{Wang2020}). 
Relying on the geometric and dynamical BLR model, 
we can extract information from RM 
and SA data to measure the cosmological distance.
However, up to now, RM campaigns measure the region of optical emission line (mainly used H$\beta$)
while the SA campaigns measure the NIR emission lines (used Pa$\alpha$ in \citealt{Gravity2018} and \citealt{Wang2020}).
Since different emission lines come from different areas of 
BLR (\citealt{Clavel1991, Dietrich1995, Kollatschny2003}), 
if RM and SA use the same emission line, 
the systematic error (mentioned in \citealt{Wang2020}) of distance measurement in joint analysis could be significantly reduced.
It is necessary to carry out SA observations in optical bands or RM observations in infrared bands.
Considering the extinction and jitter of the Earth's atmosphere, the interference process in space shows advantages in long exposure.
The space-based optical interferometer will be the future direction.

As a necessary step, we aim at 
studying the uncertainty of distance measurement 
caused by basic observational parameters of the space-based optical interferometer in this paper.
Structure of this paper is as follows. 
In section 2 we describe the geometric and dynamical model of the BLR adopted.  
In section 3 we simulate the expected spectroastrometric signals and describe the analysis method in section 4. 
Results are provided in section 5, and we show our summary and discussionsin the last section.

\section{Parameterized BLR model}           
\label{sect:BLRmodel}

In the optical band, AGN spectrum consists of continuum emission and line emission.
The continuum emission is considered to originate chiefly from the accretion disk.
For nearby galaxies, continuum emission from the host galaxy could also not been ignored.
For simplicity, 
we assume the region produced continuum emission 
has an axis symmetrical structure
and the photocenter of continuum emission locates at the black hole.
The broad emission lines are considered be generated from photoionization of BLR clouds 
by the UV continuum photons from the accretion disk.
The emission line used in our simulation is H$\beta$, 
which is a strong optical emission line 
and widely used for RM observation to estimate the liner size of BLR. 
The continuum underneath emission lines is often assumed to have a linear form.
This hypothesis is widely used in RM to calculate the integral flux of H$\beta$,
where continuous spectrum is required to be subtracted off (\citealt{Du2014}; \citealt{Huang2019}). In order to estimate the photocenter position of broad H$\beta$ emission line and simulate the SA observations, 
we use a simple BLR model that is characterized by a flat disk with circular Kepler rotation.

In recent years, thanks to many studies combining RM observation and parameterized BLR model,
we have gained a better understanding of BLR physics 
(\citealt{Pancoast2014a, Pancoast2014b, Li2013, Li2018, Grier2017a,  Williams2018}).
In this section we give an overview of the parameterized model we used for BLR. 
It consists of many isolated BLR clouds, 
which are model as a large number of non-interacting point particles. 
These particles reprocess the continuum photons originated from the accretion disk 
into the emission line photons instantaneously. 
The wavelength of the emission line photons 
is determined by the particles's velocity 
and the time lag for the response 
is determined by their position.
For simplicity, the accretion disk is regarded as a point-like geometry
so that the UV ionizing continuum is isotropic 
and the flux of continuum falls off with the square of the distance.

\subsection{Geometry}

In our simulation, 
the radial distribution of the point particles is described 
as a shifted $\Gamma-$distribution. 
The distance $r$ of a point particle from the black hole 
is given by 
\begin{equation}
r = \Gamma_0\,(1-F)\beta^2\,R_{\rm {BLR}} + R_s + F\,R_{\rm {BLR}}.
\end{equation}
Here $\Gamma_0 =p(x|1/\beta^2, 1)$ is drawn randomly from a Gamma distribution
\begin{equation}
p(x|\alpha,\theta)=\frac{x^{\alpha-1}{\rm exp}(-{x}/{\theta})}{\theta^\alpha\Gamma(\alpha)},
\end{equation}
the point particle is then shifted radially by a Schwarzschild radius
 $R_s = 2GM_{\rm {BH} }/c^2$, 
plus a minimum radius $R_{\rm{min}}$ of BLR. 
Here $R_{\rm{BLR}}$ is the mean radius,
$F =R_{\rm{min}}/R_{\rm{BLR}}$ is a fraction of the minimum radius to the mean radius. 
Within $R_{\rm{min}}$, the line-emitting particles are not allowed to exist.
$\beta$ is the shape parameter of the radial distribution: small values of $\beta$ mean narrow normal distributions while large values mean exponential distributions. 
We then define a half-opening angle $\theta_{\rm o}$ for the overall geometry 
and the spatial distribution of point particles. 
Here $\theta_{\rm o} = 0^{\circ}$ defines a thin disk (ring) 
while $\theta_{\rm o} = 90^{\circ}$ represents a spherical distribution. 
An observer views the BLR from an inclination angle $\theta_{\rm i}$,
where $\theta_{\rm i} = 0^{\circ}$ corresponds to a face-on orientation.

For each point particle, the emission is weighted by a parameter $\omega(\phi)$,
\begin{equation}
\omega(\phi)=\frac{1}{2} + k \rm\ cos(\phi),
\end{equation}
where $k$ is a free parameter between -0.5 and 0.5. 
When $k = 0.5$, the inner side of BLR is contributing more line emission.
Here $\phi$ is the angle between the observer's line of sight to the black hole 
and the point particle's line to the black hole.
The particles could be clustered, $\theta$ is the angle of a point particle from the disk:
\begin{equation}
\theta = \rm {arccos}\left( cos\,\theta_{\rm o} + (1-\rm{cos\,\theta_{\rm o}})U^{\gamma} \right), 
\end{equation}
where $U$ is a random number drawn uniformly between 0 and 1, $\gamma$ is a free parameter drawn uniformly between 1 and 5. When $\gamma = 1$, the point particles are clustered towards the disk. 

We also adopt a usual assumption that BLR clouds have the same size 
and no shadowing among them (\citealt{Li2013}).
We use the traditional linear response of the emission lines to the continuum.
For BLR that has been given geometry and clouds distribution, 
the emission-line flux $f_{\rm l}$ at time $t$ is estimated
by summing over the emissions from all the clouds: 
\begin{equation}
f_{\rm l} = A \sum_{ i} \omega_{ i} \frac{f_{\rm c}}{r_{ i}^2}
\end{equation}
Here $\omega_{ i}$ is the weight of the $i$ th clouds calculated by Equation(3), $A$ is the response coefficient and $f_c/r_{i}^2$ described the ionizing flux received by the $i$th cloud at time $t$.

\subsection{Dynamics}
We assume a flattened disk-like BLR with Keplerian rotation. 
Due to the gravity of the central black hole, the point particles rotate in a circular orbit.
The Keplerian velocities $V_{\rm K}$ of the point particles are drawn from a distribution that depends upon the black hole mass $M_{\rm{BH}}$ 
and the  distance $r$, which could be estimated as:
\begin{equation}
V_{\rm K} = \left( \frac{GM_{\rm{BH}}}{r} \right)^{1/2}.
\end{equation}
For a thick Keplerian disk with a half opening angle $\theta_{\rm o}$, 
there will be an angle $\theta$ between the orbital plane and the equatorial plane.
Thus, the LOS velocities $v_{\rm {line}}$ are related to the Keplerian velocities, 
half-opening angle and the inclination angle.
For cloud particles with an emitted wavelength $\lambda_{\rm {emit}}$, 
considering the relativistic effect, Doppler broadening and gravitational redshift 
which will affect the emission line profile, the observed wavelength $\lambda_{\rm {obs}}$ can be written as:
\begin{equation}
\lambda_{\rm{obs}} = \left( 1+\frac{v_{\rm{line}}}{c}\right) \left(1-\frac{V_{\rm{K}}^2}{c^2}\right)^{-1/2}\left(1-\frac{R_{\rm S}^2}{r}\right)^{-1/2}\lambda_{\rm{emit}}.
\end{equation}

In order to explore the influence of basic observational parameters on the uncertainty of distance measurement, 
we adopt the same set of BLR model parameters as shown in \cite{Wang2020}, which has been summarized in Table \ref{table:blrpara}.

\begin{table}
\bc
\begin{minipage}[]{100mm}
\caption[]{Parameters used in the BLR model \label{table:blrpara}}\end{minipage}
\setlength{\tabcolsep}{5pt}
\small
 \begin{tabular}{llll}
  \hline\noalign{\smallskip}
Parameters & meanings & value & Prior ranges \\
  \hline\noalign{\smallskip}
F & fractional inner radius of BLR &  0.49 & [0,1]      \\
$\beta$ & radial distribution of BLR clouds & 1.09 & [0,4]\\
$\theta_{\rm{o}}(^{\circ})$ & half opening angle of the BLR & 39.96 & [0,90]\\
$\theta_{\rm {i}}(^{\circ})$ & inclination angle of the BLR & 8.41 & [0,90]\\
PA$(^{\circ})$ & position angules & 210.99 & [0,520]\\
$R_{\rm {BLR}}$(ltd) & averaged linear sizes & 184.17 & [1,10$^3$]\\
$M_{\rm{BH}}$(10$^8M_{\rm{sun}}$) & supermassive black hole mass & 5.78 & [10$^{-2}$,10]\\
$D_{\rm{A}}$(Mpc) & absolute angular distance & 551.50 & [10,10$^{4}$]\\
  \noalign{\smallskip}\hline
\end{tabular}
\ec
\tablecomments{0.86\textwidth}{The parameters were used in BLR model and the values obtained from \cite{Wang2020}.}
\end{table}

\section{the spectroastrometric signal of the BLR}
\label{sec:signal}

The SA method (see \citealt{Bailey1998, Rakshit2015}) 
shows advantages in providing information on the spatial structure of the object,
especially when the global angular size $\Lambda$ of a non-resolved source 
smaller than the interferometer limit $\lambda/{ B}$,
here $\lambda$ is the observed wavelength 
and ${ B}$ is the length of interferometer baseline.
By measuring the photocenters of different wavelength channels, 
this method has been used to observe AGNs (\citealt{Petrov2001, Marconi2003, Gnerucci2010, Gnerucci2011a, Rakshit2015}) in near-infrared 
and successfully constrains the size of BLR for 3C273 (\citealt{Gravity2018, Wang2020} ). 
In this section we give a description of how we simulate SA signal.


For a given object, the observed interferometric phase $\Phi$ at wavelength $\lambda$ can be writren as 
\begin{equation}
\Phi(\lambda) = \Phi_{\ast} (\lambda) + \Phi_{\rm t}(\lambda), 
\end{equation}
where $\Phi_{\ast}(\lambda) $ is the phase contributed from real astrophysical signal 
and $\Phi_{\rm t}(\lambda) $ is the phase contributed from 
the atmosphere, telescopes, interferometric delay lines 
and the various instrumental parts down to the detector(see \citealt{Vannier2006}). 
\cite{Petrov1989} shows that, for an interferometer with a baseline $\bm B$, the interferometric phase for the object is

\begin{equation}
\Phi_*(\lambda) = -2 \pi \frac{\bm{B}}{\lambda} \cdot \bm \epsilon(\lambda).
\label{eq:dp}
\end{equation}
Here $\bm \epsilon(\lambda)$ is the photocenter of the source at wavelength $\lambda$ 
and could be written as
\begin{equation}
\bm \epsilon(\lambda) = \frac{\int\int \bm r O(\bm r,\lambda)d^2\bm r }{\int\int O(\bm r,\lambda)d^2\bm r },
\label{eq:epsilon}
\end{equation}
$O(\bm r,\lambda)$ is the brightness distribution of the source. 
If defining the fraction of the flux of the emission line to total as $f_l(\lambda) = F_l(\lambda)/F_{\rm{tot}}(\lambda)$, where $F_l(\lambda)$ is the flux of the emission line at wavelength $\lambda$ and $F_{\rm{tot}}(\lambda)$ is total flux at wavelength $\lambda$, we have
$
\bm \epsilon(\lambda) = f_l(\lambda)\bm \epsilon_l(\lambda)
$,
$\bm r$ is the vectors to the central black hole and
\begin{equation}
\bm \epsilon_l(\lambda) =  \frac{\int\int \bm r  O_l d^2\bm r }{\int\int O_ld^2\bm r }.
\label{eq:epsilonl}
\end{equation}
We used the bold letters mark the vectors $\bm{B}$, $\bm \epsilon(\lambda)$ and $\bm r$.

For space-based interferometers, 
the phase contributed from the atmosphere could be ignored but the phase contributed from instrument could not.
In principle, if we add a spatial filtering before the beam recombination, 
the optical effects left after this spatial filtering 
only include differences in intensity between beams and an OPD (optical path difference).
The previous one can be calibrated by photometric channels. 
So only the phase shift
caused by the OPD is taken into consideration
in our simulation.
The OPD has two origins:
one originates exclusively from piston 
and chromatic dispersion along the path 
before the filtering, 
another originates from wavefront corrugations
induced by imperfect adaptive optics.(\citealt{Tubbs2005}; \citealt{Vannier2006}).
For a given spectral channel, the phase shift caused by OPD is 
\begin{equation}
\Phi_{\rm{OPD}}(\lambda) = (2\pi/\lambda){\rm{OPD}}.
\label{eq:phiopd}
\end{equation}
Hence, the differential phase between a channel $\lambda_{\rm i}$ and 
the reference channel $\lambda_{\rm r}$ will be:
\begin{equation}
\Delta\Phi(\lambda_{i},\lambda_{\rm r}) = \left[\Phi_{\ast}(\lambda_i) - \Phi_\ast(\lambda_{\rm r}) \right] + \left[\Phi_{\rm{OPD}}(\lambda_i) - \Phi_{\rm{OPD}}(\lambda_{\rm r}) \right].
\label{eq:dpclambda}
\end{equation}
By inserting Equation(\ref{eq:dp}), Equation(\ref{eq:epsilon}) and Equation(\ref{eq:phiopd}) into Equation(\ref{eq:dpclambda}), we have

                                                                                                                                                                                      \begin{equation}
\Delta \Phi(\lambda_i, \lambda_r) 
= -2 \pi {\bm B} \cdot 
[ f_l(\lambda_i)\frac{{\bm \epsilon_i} (\lambda_i)}{\lambda_i} - f_l(\lambda_r) \frac{{\bm \epsilon_r} (\lambda_r)}{\lambda_r}]
+2\pi \cdot{\mathrm {OPD}} \cdot (\frac{1}{\lambda_i}-\frac{1}{\lambda_r}).
\label{eq:dpcmodel}
\end{equation}

If the BLR model is specified,                                                                                                                                                                                                                       the surface brightness distribution of the regions                                                                                                                                                                                                        is mainly related to the total number of collecting photons,
which is linearly dependent on the number of telescopes ($n_{\rm T}$), 
the overall quantum efficiency ($\eta$), 
the collecting area of telescopes $S$ and the exposure time $t$.
For each wavelength channel, 
the observed photocenter is further determined by the choice of spectral resolution $R$. 
If we also know the configuration of the projected baseline $\bm B$, 
we could predict the actual phase signal by Equation(\ref{eq:dp}).
By giving the limit OPD (rms) control errors during an observation,
using Equation(\ref{eq:dpcmodel}), 
we could obtain wavelength-dependent differential phase 
afterwards differential phase curves (DPCs).
For simplicity, the equivalent diameter ($D$) is used to estimate the number of photons for unit time in our simulation,
$S = \pi (D/2)^2$.

\begin{figure} 
   \centering
   \includegraphics[width=15.0cm, angle=0]{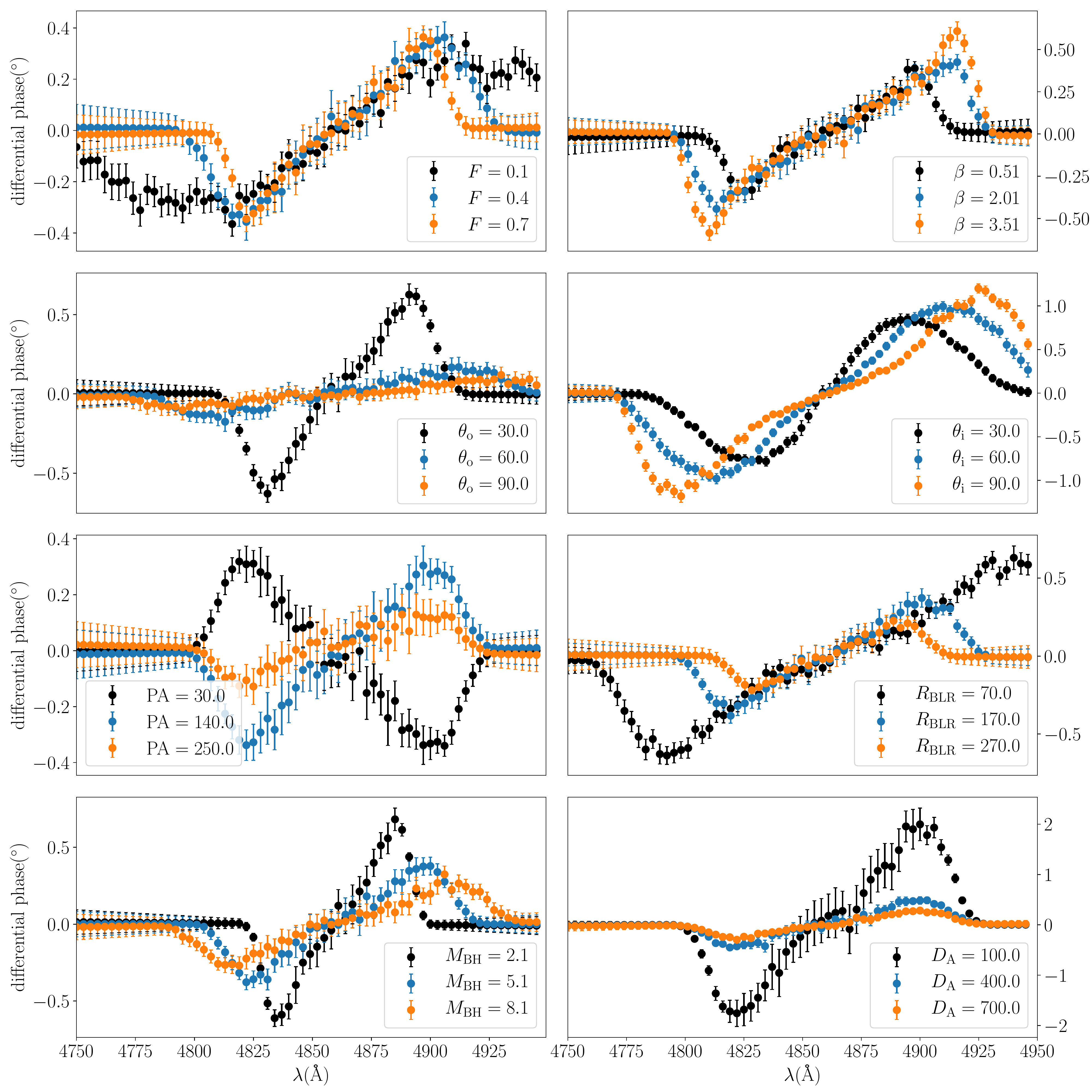}
   \caption{Dependence of DPCs on eight parameters. The parameters for interferometer are fixed at the values listed in Table\ref{table:interferometry}. 
When varying one parameter, other parameters are fixed at the values listed in Table\ref{table:blrpara}. The parameters and their values are shown in legend. The units of $\theta_{\rm o}, \theta_{\rm i}, \rm{PA} $ are degrees. The unit for $R_{\rm{BLR}}$, $M_{\rm{BH}}$ and $D_{\rm{A}}$ are lt-d, $10^8M_{\rm{sun}}$ and Mpc, respectively.
} 
   \label{Fig:DPC_PARA}
\end{figure}

In our simulation, uncertainty of the signal consists of two parts,
one part comes from Poisson noise 
and the other caused by the uncertainty of OPD control.
Each set of scientific exposures is assumed 
to be the average over $n_{\rm{obs}}$ (the number of observations) frames of t-s integration (NDIT = $n_{\rm{obs}}$ and DIT = $t$ s).
We use the average value of $n_{\rm{obs}}$ observations at wavelength $\lambda$ as
the differential phase  $\Delta\Phi(\lambda)$ 
and the dispersion of $n_{\rm{obs}}$ observations as the error.
We show the dependence of mock DPCs on eight parameters used in BLR model in Figure\ref{Fig:DPC_PARA}. 

\section{analysis}
\label{sec:analysis}

In this section, we describe the analytical method of obtaining BLR model parameters from mock data.
The joint analysis proposed by \cite{Wang2020} uses data of three parts: 
light curves of optical continuum and emission lines,  
DPCs and the profiles of the emission lines. 
The light curves and emission profiles depend on optical spectral observation, 
which has been a very mature technology, 
so we assume that the quality of the light curves 
and the profiles of the emission lines are good enough.
Thus, the uncertainty of the joint analysis depends mainly on the data quality of DPCs.


As mentioned above, we generate the mock DPCs from the Equation(\ref{eq:dpcmodel}).
We use the DPCs to obtain the posterior distribution of the BLR model parameters.
Suppose that probability distributions for the measurement errors 
of the DPCs are Gaussian and uncorrelated, 
the likelihood function can be written as 
 \begin{equation}
 P(\mathcal{D}|\Theta) = \prod\limits_{i=1}^{N_{\rm {obs}}}\prod\limits_{j=1}^{N_{\lambda}}\frac{1}{\sqrt{2\pi\sigma^2_{\phi_{ij}}}}
 \rm {exp}\left\{-\frac{[\Phi_{\rm {obs}}-\Phi_{\rm{mod}}(\Theta)]^2}{2\sigma^2_{\phi_{ij}}}\right\}.
 \end{equation}
Here $\mathcal{D}$ represents the measured data, 
$\Theta$ represents the BLR model parameters, 
$\Phi_{\rm{obs}}$ is the interferometric phase of the emission line with the uncertainties $\sigma_{\phi_{ij}}$, 
$\Phi_{\rm{mod}(\Theta)}$ is the corresponding predicted values from the BLR model,
$N_{\rm{obs}}$ is the number of observations 
and $N_{\lambda}$ is the corresponding number of the wavelengthg bins.
According to Basyes' theorem, 
if we know the prior distribution of the model parameter, 
the posterior probability distribution for $\Theta$ should be given by 
\begin{equation}
 P(\Theta|\mathcal{D}) =\frac{P(\Theta) P(\mathcal{D}|\Theta) }{ P(\mathcal{D})},
 \end{equation}
 where $P(\mathcal{D})$ is a normalization factor.
We list the parameters of BLR model and the prior ranges in Table \ref{table:blrpara}.

EMCEE package is used to sample the parameters efficiently (\citealt{Foreman2013}), 
which is an MIT licensed pure-Python implementation 
of Goodman $\&$ Weare’s Affine Invariant Markov chain Monte Carlo (MCMC) Ensemble sampler. 
The sampling is run with 1000 ``walkers" independently 
and converges within about 2000 trials. 

\section{results}
Utilizing the BLR model described in Section \ref{sect:BLRmodel}, 
we generate mock observations 
according to a set of varying basic observational parameters ($B,D,t,R$) described 
in Section \ref{sec:signal}. 
Combined with the method mentioned in Section \ref{sec:analysis}, 
we first study the effect of inclination angle and the projected angle on the photocenter shift, 
then we study the influence of the basic observational parameters on the data quality of DPCs and the accuracy of distance measurement.

\subsection{the effect of inclination angle on the photocenter shift}

\begin{figure} 
   \centering
   \includegraphics[width=15.0cm, angle=0]{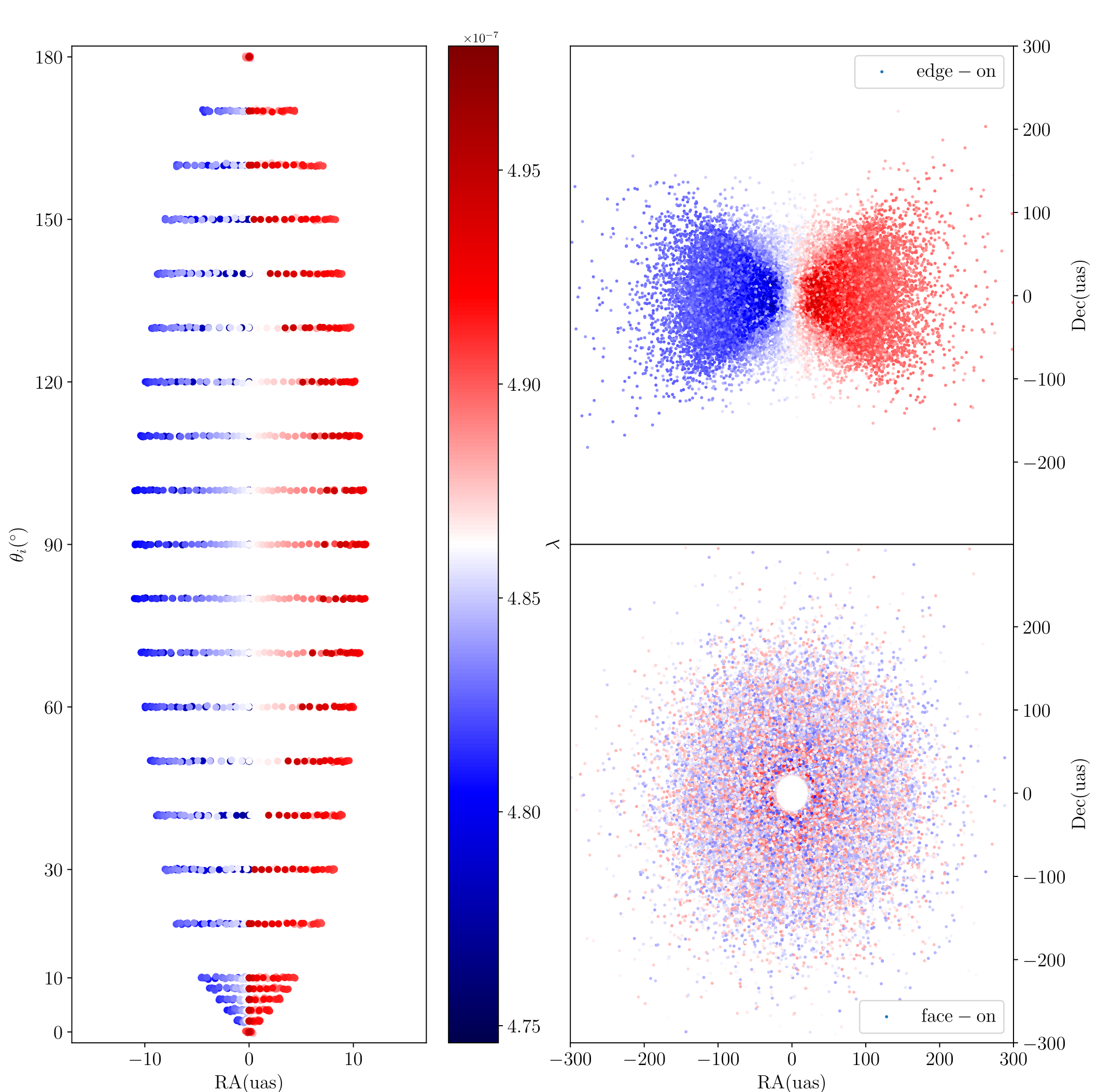}
   \caption{The effect of inclination angle on the photocenter shift.
   Left panel: the calculated distribution of photocenters that vary with the inclination angular. Upper right panel: The position distribution of photocenters at different wavelengths seen by an observer from the edge-on direction ($\theta_i = 90^{\circ}$). 
Lower right panel: The position distribution of photocenters at different wavelengths seen by an observer from the face-on direction ($\theta_i = 0^{\circ}$).} 
   \label{Fig:n_photon}
\end{figure}

The SA method can spatially resolve the velocity gradient perpendicular to LOS direction, the size of which varies with the inclination angle. Therefore, we first study how inclination angle affects the distribution of photocenters.
We used the hypothesized BLR model mentioned above. 
The parameters of model are fixed 
and listed in Table \ref{table:blrpara}, except the inclination angle. 
We vary the inclination angle 
($\theta_i$) from 0$^{\circ}$ to 180$^{\circ}$.
Then we estimate the corresponding distribution of photocenters which are shown in left panel of Figure\ref{Fig:n_photon}.
By changing the inclination angle, 
we find that for BLR with Keplerian motion, 
the more ``edge-on" the LOS is,
the larger the spatial distribution size of wavelength-dependent photoncenters. 
When viewed in a direction perpendicular to the disk plane (face-on: $\theta_i=0^{\circ}$), 
the photocenters of different wavelengths could not be distinguished (as shown in lower right panel of Figure\ref{Fig:n_photon}).
This conclusion can be understood as when viewed from angular $\theta_i$, the velocity components along LOS direction are proportional to sin $\theta_i$. 
Therefore, when viewed form the edge-on direction ($\theta_i= 90^{\circ}$),  
the velocity along LOS direction achieved maximum such that the observed wavelength gradient is the largest (shown in upper right panel and lower right panel of Figure\ref{Fig:n_photon}).  
We must emphasize that the shielding of dust torus is not taken into account in our simulation.

\subsection{the influence of projected angle on the photocenter shift}
\begin{figure} 
   \centering
   \includegraphics[width=15.0cm, angle=0]{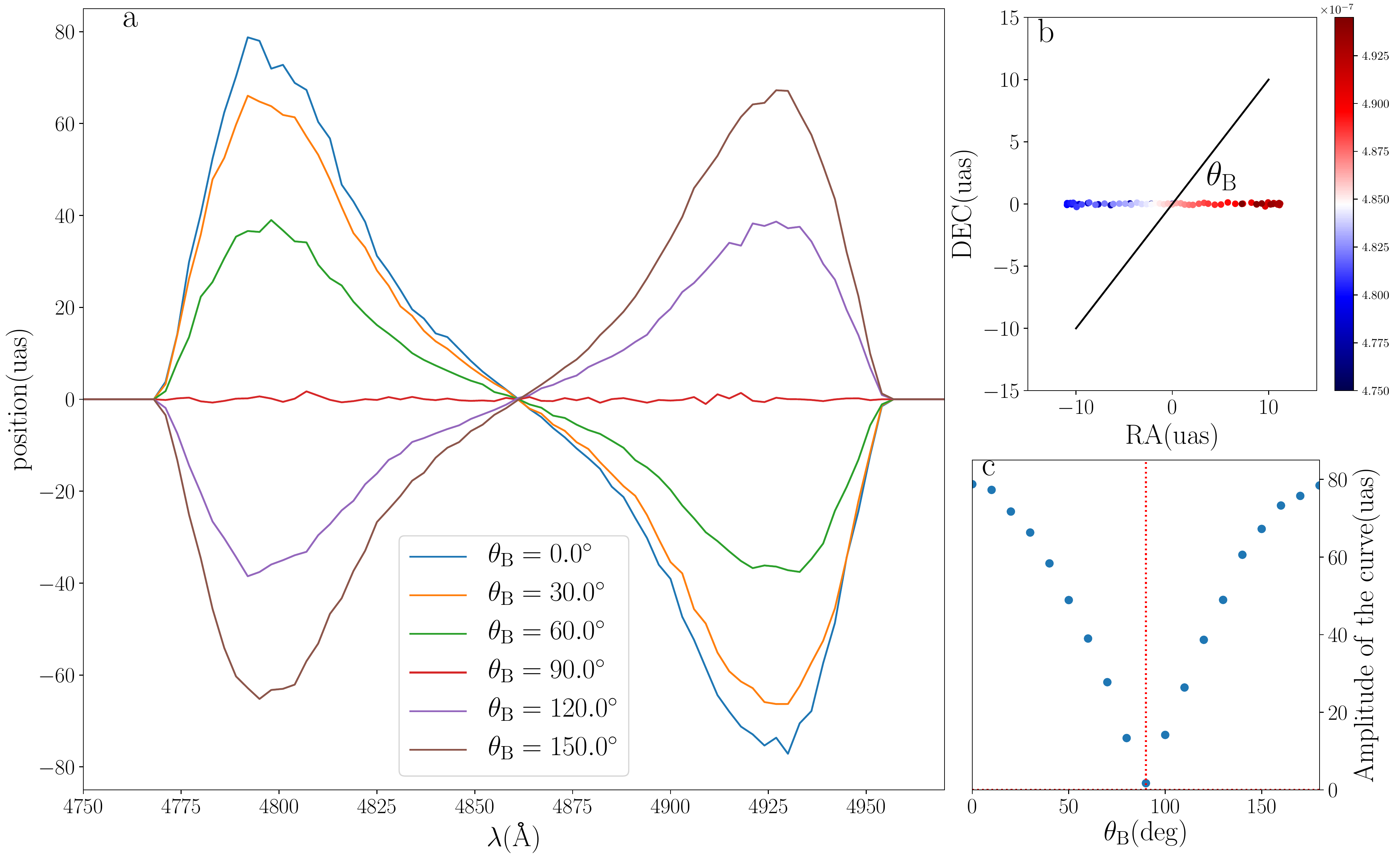}
   \caption{The influence of projected angle on the photocenter shift. Left panel: simulated position of photocenters at different wavelengths under different projected angle $\theta_{\rm B}$. Upper right panel: the black solid line shows the baseline direction, colored points represent the photocenters of different wavelengths and the $\theta_{\rm B}$ is the projected angle. 
Lower right panel: the maximum offset of photocenters vary with $\theta_{\rm B}$. 
} 
   \label{Fig:ang_amp}
\end{figure}

In fact, the angle between the baseline and the plane of BLR 
will also affect the observed distribution of photocenters.
This angle is defined as the projected angle (shown in Figure\ref{Fig:ang_amp}.b as $\theta_{\rm B}$) in our simulation.
Colored points represent the photocenters of different wavelengths, and the black solid line represents the baseline direction.
Each baseline can only measures the distribution of photocenters projected to the baseline direction, which could be indicated from Equation {\ref{eq:dp}}.  
To study the impact of the angle $\theta_{\rm B}$, we use the same BLR model as mentioned above 
and fix the $\theta_i = 90^{\circ}$, 
then vary the angle $\theta_{\rm B}$
to measure the displacement of photocenters at different wavelengths. 
Results are shown in Figure \ref{Fig:ang_amp}.a and the relation between the maximum offset of photocenters and $\theta_{\rm B}$ 
is shown in Figure\ref{Fig:ang_amp}.c.
We see that the maximum offset of photocenters are proportional to the absolute value of cos $\theta_{\rm B}$.
When the baseline is perpendicular to the plane, 
the photocenters of different wavelengths along the baseline direction is indistinguishable. 

In actual observations, the direction of the BLR plane (so the projected angle $\theta_{\rm B}$) is unknown to the observer.
To ensure that the mock data contains information about spatial position, we adopted an observation strategy of rotating the baseline.
In our simulation, we rotate the baseline regularly with a cadence of 10${^{\circ}}$.
Thus we get 19 mock data for each set of observational parameters.

\begin{table}
\bc
\begin{minipage}[]{100mm}
\caption[]{Parameters used in the interferometry model \label{table:interferometry}}\end{minipage}
\setlength{\tabcolsep}{5pt}
\small
 \begin{tabular}{llll}
  \hline\noalign{\smallskip}
Parameters & meanings & value & range \\
  \hline\noalign{\smallskip}
$\bm B$(m) & the interferometer baseline  &  100 &  [20,1000]  \\
D(cm) & the diameter of the telescope & 100 & [20,1000]\\
t(s) & the time of exposure & 2000 & [200,20000]\\
R & the resolution of spectrum & 1000 & [250,10000]\\
$n_{\rm{obs}}$ & the number of the observations & 30 & - \\
${\rm{OPD}}$(nm) & the accuracy of OPD control & 5 & -\\
  \noalign{\smallskip}\hline
\end{tabular}
\ec
\tablecomments{0.86\textwidth}{The parameters were used in interferometer model and the range of the change. In our simulation, the ${\rm OPD}$ is a random value 
chosen from a Gaussian distribution 
with dispersion 5nm (rms), 
which equal to a absolute phase shift
of $\sim$ 0.06 rad (3.7$^{\circ}$) in 4861 $\rm \AA$.}
\end{table}

\subsection{the influence of basic observational parameters on data quality}

Furthermore, we study the effects of the basic observational parameters
by varying one parameter at a time (the ranges of variations are summarized in Table {\ref{table:interferometry}}).
In addition, the projected angle $\theta_{\rm B}$ was fixed at $\theta_{\rm B} =0 ^{\circ}$.

\begin{figure} 
   \centering
   \begin{minipage}{1.0\textwidth}
   \includegraphics[width=15.0cm, angle=0]{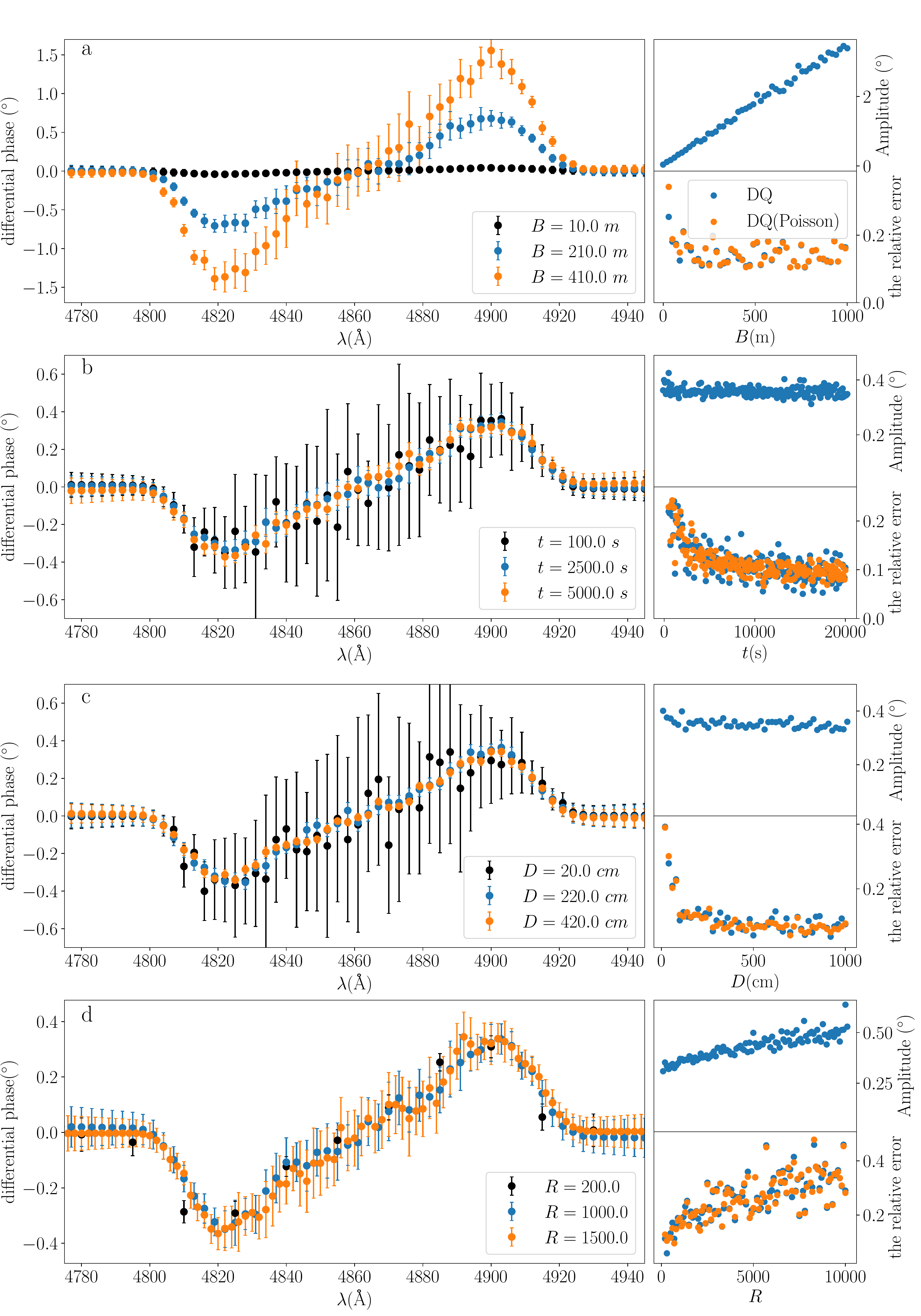}
   \end{minipage}
   \caption{{ Comparison of differential phase curves and data quality under different parameters.  When varying one parameter, other parameters are fixed at the values listed in Table 2. ({\it panel a}: baseline length B; {\it panel b}: total integration time $t$; {\it panel c}: equivalent diameter $D$; {\it panel d}: spectral resolution $R$). {\it Left column}: Comparison of simulated differential phase curves with error under different values of the parameter. {\it Right panel}: data quality (DQ) for DPCs differing only the parameter. ({\it upper}: Amplitudes of the curves; {\it lower}: the relative error of DPCs, blue for total error and orange for Poisson error)}. }
   \label{Fig:phase}
\end{figure}

In order to quantify the impact of basic observational parameters mentioned above, we define two metrics to describe the data quality: the largest amplitude 
and the relative error of DPCs. 
The largest amplitude of DPCs is defined as the maximum absolute amplitude of a set of DPCs.
The relative error of DPCs is quantified as the ratio of phase error to phase value where phase is the largest amplitude of DPCs.
We use the relative Poisson error of DPCs, 
which is defined as the ratio of the Poisson error to 
the largest amplitude of DPCs,
to quality the impact of Poisson noise.
The results are shown in Figure \ref{Fig:phase}.

The impact of baseline length on data quality is shown in Figure\ref{Fig:phase}.a.
We show three DPCs generated under three different baseline lengths in the left panel.
These DPCs show obvious S-shape. 
The value of the differential phase increases as the baseline gets longer, so does the absolute error.
We find that the largest amplitude of DPCs 
increases linearly with the baseline length (shown in upper right panel), which has been implied from the Equation \ref{eq:dp}.
As the baseline gets longer, the relative error tends to be smaller (shown in bottom right panel).
We also find that the Poisson error becoming dominating while the baseline is getting longer.

The impact of the exposure time is shown in Figure\ref{Fig:phase}.b.
We also show three DPCs generated under three different exposure times in the left panel.
It is easy to see that as the exposure time increases, the absolute error of DPCs decrease significantly,
but the value of the differential phase remains essentially the same.
As shown in the upper right panel, the largest amplitude of DPCs does not change with exposure time.
However, relative error and relative Poisson error of the DPCs decrease as the exposure time increases. 

The influence of the equivalent diameter is basically the same as that of exposure time, and the results are shown in the Figure\ref{Fig:phase}.c.
As the equivalent diameter becomes larger, absolute error, relative error and relative Poisson error of the DPCs are significantly reduced, but the largest amplitude of DPCs remains the same.
For our simulation, we find that by changing the equivalent diameter, relative error of the DPCs can be reduced to less than 5$\%$, which is more effective than increasing the exposure time to roughly 10$\%$.

The results of how the spectral resolution impacts the DPCs are shown in Figure\ref{Fig:phase}.d. 
As the resolution increases, 
the number of data points on the DPC
increases significantly (shown in left panel), while the largest amplitude of DPCs increases slightly (as shown in upper right panel) and the relative error of DPCs becomes larger (as shown in bottom right panel).
This conclusion can be understood as the larger the spectral resolution, the more the wavelength channels, but the less the number of photons in each wavelength channel.


\subsection{ the effect of basic observational parameters on the accuracy of distance measurement}

\begin{figure} 
   \centering
   \includegraphics[width=15.0cm, angle=0]{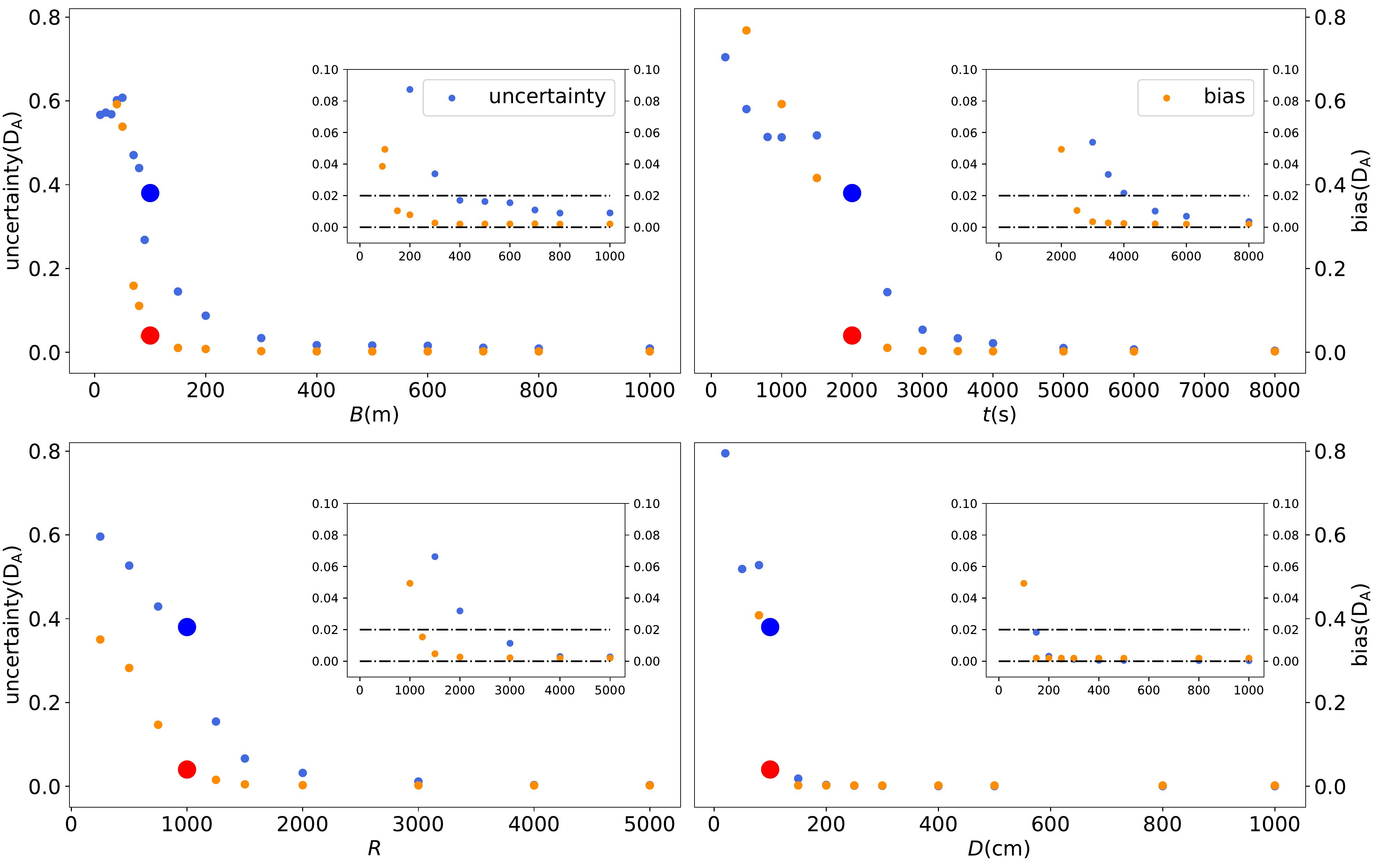}
   \caption{The relative uncertainty (blue points) and bias (orange points) of distance measurement change with each parameter. Upper left panel for baseline length. Upper right panel for exposure time. Lower left panel for spectral resolution and lower right panel for equivalent diameter. The large point is the result corresponding to the typical value listed in Table \ref{table:interferometry}.   
    } 
   \label{fig:B_DA}
\end{figure}

In order to quantify the accuracy of distance measurement, 
we define the median value of the posterior distribution 
as the best value for distance inferred from MCMC,
the 16$\%$ and 84$\%$ quantiles as the lower and the upper bounds.                                                                        
The relative uncertainty of the distance measurement is defined 
as the ratio of the error (mean of the lower and upper bounds) 
to the best value mentioned above.
The relative bias is defined as the fraction of the bias (difference between the best value and the input value) to the input value.
We simulate the actual DPCs using the BLR model mentioned in Section \ref{sect:BLRmodel} 
and the basic observational parameters described in Section \ref{sec:signal}. 
Then we used MCMC method described in Section \ref{sec:analysis} 
to obtain probability density distributions of BLR model parameters.
In the Appendix \ref{sec:appendix}, we list the simulated DPCs (Figure \ref{fig:appendix1}) generated by using the instrument parameters listed in Table \ref{table:interferometry}, as well as the probability density distributions (Figure \ref{fig:appendix3}) of model parameters.
We obtain that the relative uncertainty (blue points) and bias (orange points) of distance measurements ($D_{\rm{A}}$)
at different baseline lengths, exposure times, equivalent diameters and spectral resolutions, respectively.
The results are shown in Figure \ref{fig:B_DA}.

We see that as the parameter values increase, 
the relative uncertainty of distance measurement ($D_{\rm{A}}$) can be reduced to 2$\%$.
There is an optimum value for each parameter.
Above this optimum value, the accuracy of $D_{\rm{A}}$ will improve slowly with the parameter value.  
Extending the baseline can proportionally amplify the difference phases and their Poisson errors, 
which means that the influence of OPD errors can be reduced during the fitting process. 
Therefore, extending the baseline can improve the accuracy of distance measurement.
Either increasing the exposure time or the equivalent diameter results in improving the accuracy of distance measurement,
since the more photons collected, the smaller the absolute Poisson error is. 
Interestingly, the larger the resolution is, the larger the relative error of DPCs, 
but the smaller the uncertainty of $D_{\rm{A}}$ will be.
This is because as the resolution increases, 
the number of photons in each wavelength interval decreases
and thus the Poisson error of the difference phase increases, 
meanwhile the amount of data on the DPC increases. 
As expected, the relative bias of $D_{\rm{A}}$ becomes smaller as the parameter becomes larger, even more sensitive than the relative error.


\subsection{ the effect of data quality on the accuracy of distance measurement}

\begin{figure} 
   \centering
   \begin{minipage}{1.0\textwidth}
   \includegraphics[width=15.0cm, angle=0]{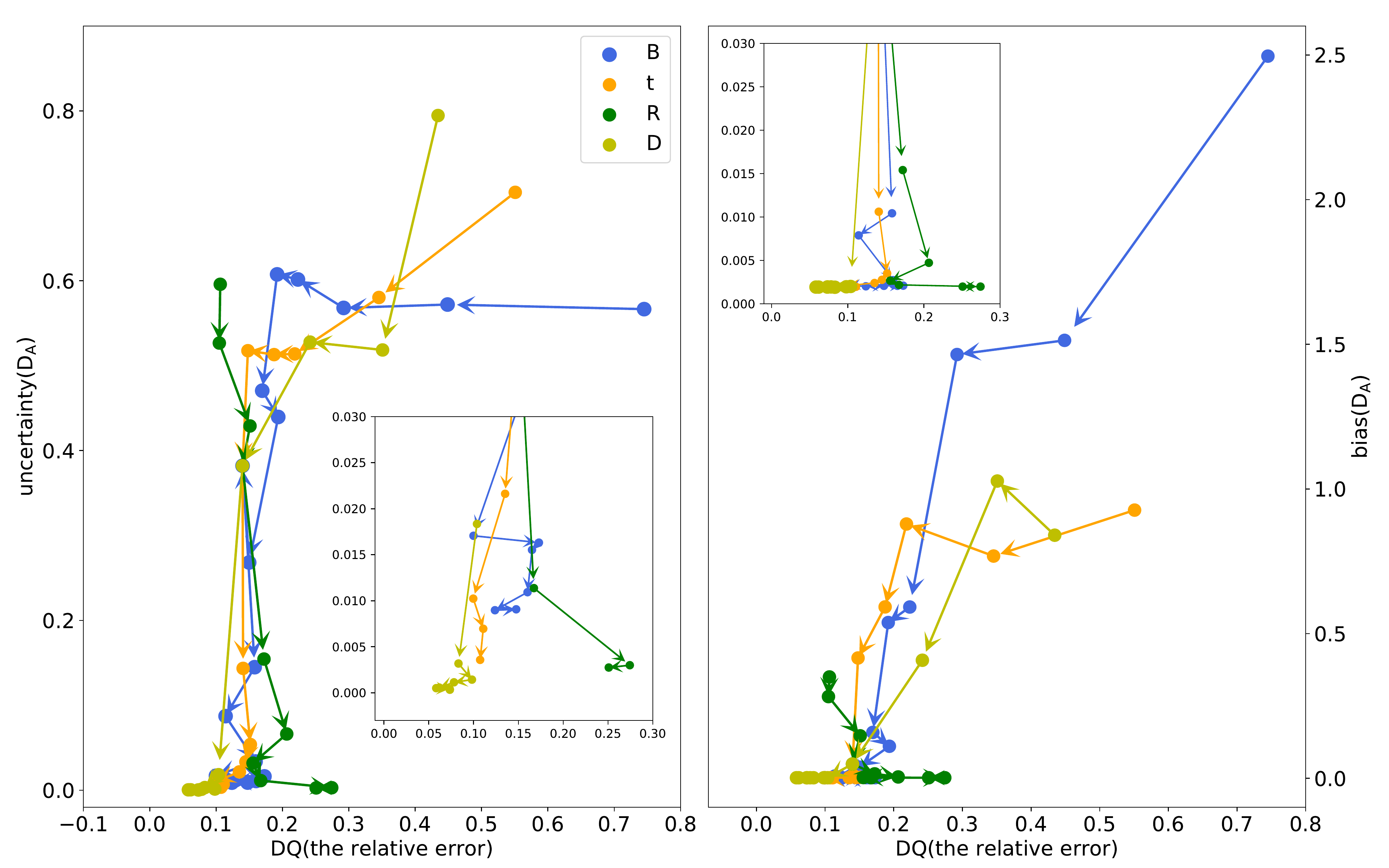}
   \end{minipage}

   \caption{The relationshift between the relative error of DPCs and the accuracy of distance measurement.
   Left panel: The relative uncertainty of distance measurement and the relative error of DPCs under different parameters.
Right panel: The relative bias of distance measurement and the relative error of DPCs under different parameters. The arrow direction is the direction in which the parameters increase.   
 } 
   \label{fig:DA_DQ}
   
\end{figure}

We also study how the uncertainty of distance measurement depends on the data quality of DPCs.
Figure \ref{fig:DA_DQ} shows the relationship resulting from 
our simulation.
We see a obvious turnaround near the data quality around 20$\%$.
Only when the data quality is better than this turnaround point, 
the accuracy of $D_{\rm{A}}$ is highly sensitive to the data quality. 
Different colors in the Figure \ref{fig:DA_DQ} 
represent different observational parameters.
The direction of the arrow between data points 
represents that parameters become larger and larger. 

As the baseline increases (blue points), 
data quality gets better first, 
but there is no effect 
on the accuracy of distance measurement. 
When the relative error of DPC reaches better than 20$\%$, 
extending the baseline can significantly improve the accuracy of $D_{\rm{A}}$,
although the data quality remains basically the same. 
Similar results are found by increasing $t$ and $D$.
However, increasing the spectral resolution (green points) gives more complex results.
As the spectral resolution increases, data quality slowly deteriorates,
but the accuracy of $D_{\rm{A}}$ is increasing. 
This may be due to fact that the larger the spectral resolution, 
the more data points on the DPCs,
but the fewer photons there are in each wavelength interval. In actual observations, when DPCs with higher spectral resolution but lower data quality are obtained, it is customary to rebin the DPCs to reduce the resolution but increase the data quality. 
However, using DPCs that have not been rebinned is a better option for improving the accuracy of $D_{\rm A}$.

\section{summary and discussion}

In this paper, we have studied the application of the SA method using space-based  optical interferometer to measure cosmological distance of quasars, and we have discussed how the basic observational parameters affect the accuracy of distance measurements. Our main results are shown as: 
\begin{itemize}
	\item[1)]{For the same target, the observed spatial distribution of the photocenters at different wavelengths will be affected by inclination angle $\theta_{\rm i}$ and the projected angle $\theta_{\rm B}$.}
	\item[2)] Four basic observational parameters will affect the data quality of DPCs. As the baseline gets longer, the value of differential phase increase, so does the absolute error. 
When the exposure time or equivalent diameter increases, the absolute error, the relative error and the relative Poisson error decrease significantly.
However, the larger the spectral resolution is, 
the larger the amplitude and relative error of DPCs are.  
	\item[3)] Four basic observational parameters will affect the accuracy of distance measurement. Extending the baseline will amplify the difference phases and the Poisson errors but will reduce the impact of OPD errors during the fitting process, so the uncertainty and bias of distance measurement both decrease.
Increasing the exposure time or effective aperture can reduce the Poisson error due to the increase in the number of photons, thus improving the accuracy of distance measurement. 
With the increase of spectral resolution, the poisson error increases due to the decrease of photon number in each wavelength channel, but the number of data on DPC increases, so although the relative error of DPC becomes larger, the uncertainty of distance measurement becomes smaller.

\end{itemize}

By assuming a parameterized BLR model and specifying the value of parameters,
we obtained the surface brightness distribution
of the BLR and continuum regions, 
then calculated the photocenters at different wavelengths.
Combined with the basic observational parameters, we conduct an extensive set of simulations to get simulated DPCs.
And then we used MCMC method to obtain the posterior probability distributions of BLR model parameters from the mock data. 
In our simulation, 
we can eventually limit the relative uncertainty of data to 10-20$\%$ by extending baseline. 
The exposure time eventually limits the relative uncertainty to 5-15$\%$ while the equivalent diameter can limit to 5-10$\%$. 
And the results show that SA method makes sense for distance measurement only if the relative error of DPCs is less than 20$\%$.
In this case, further increases in baseline length, exposure time and equivalent diameter will not greatly reduce the relative error of DPCs, but the uncertainty of distance measurement can be sharply reduced. 
However, it should be emphasized that these thresholds of the relative error are only valid for our simulation, since they depend on the specific set of BLR model and simulated observations.
The simplest model for BLR is used in our simulation.
Systematic errors, which caused by complicated structures and kinematics of BLR, such as non-disc structure, non-Keplerian kinematics and the disordered motion of these clouds, are not taken into account.
So the relative error of DPCs would be underestimate during simulation then the accuracy of model parameters would be overestimate. 
On the other hand, we did not consider the uncertainty of RM data in the fitting process, which also resulted in overestimation of the accuracy of model parameters.

It also needs to be emphasized that both RM and SA are required to measure $D_{\rm A}$ in reality.
The reason that we can use SA to determine $R_{\rm{BLR}}$ (without using the light curves from RM) is because a specific BLR model is used in our simulation to simulate DPCs, and then the same BLR model is used to perform the DPCs fitting to retrieve all the model parameters. 
This is also why the degeneracy among the eight parameters is so weak. 
However, in reality the geometric distribution, dynamics, and reprocessing coefficient of BLR clouds are much more complex, so it is not possible to have a perfect BLR model for DPCs fitting, hence it still requires RM to measure $R_{\rm{BLR}}$.

\newpage
\appendix
\section{mock data}
\label{sec:appendix}

\begin{figure}
   \centering
   \begin{minipage}{1.0\textwidth}
   \includegraphics[width=15.0cm, angle=0]{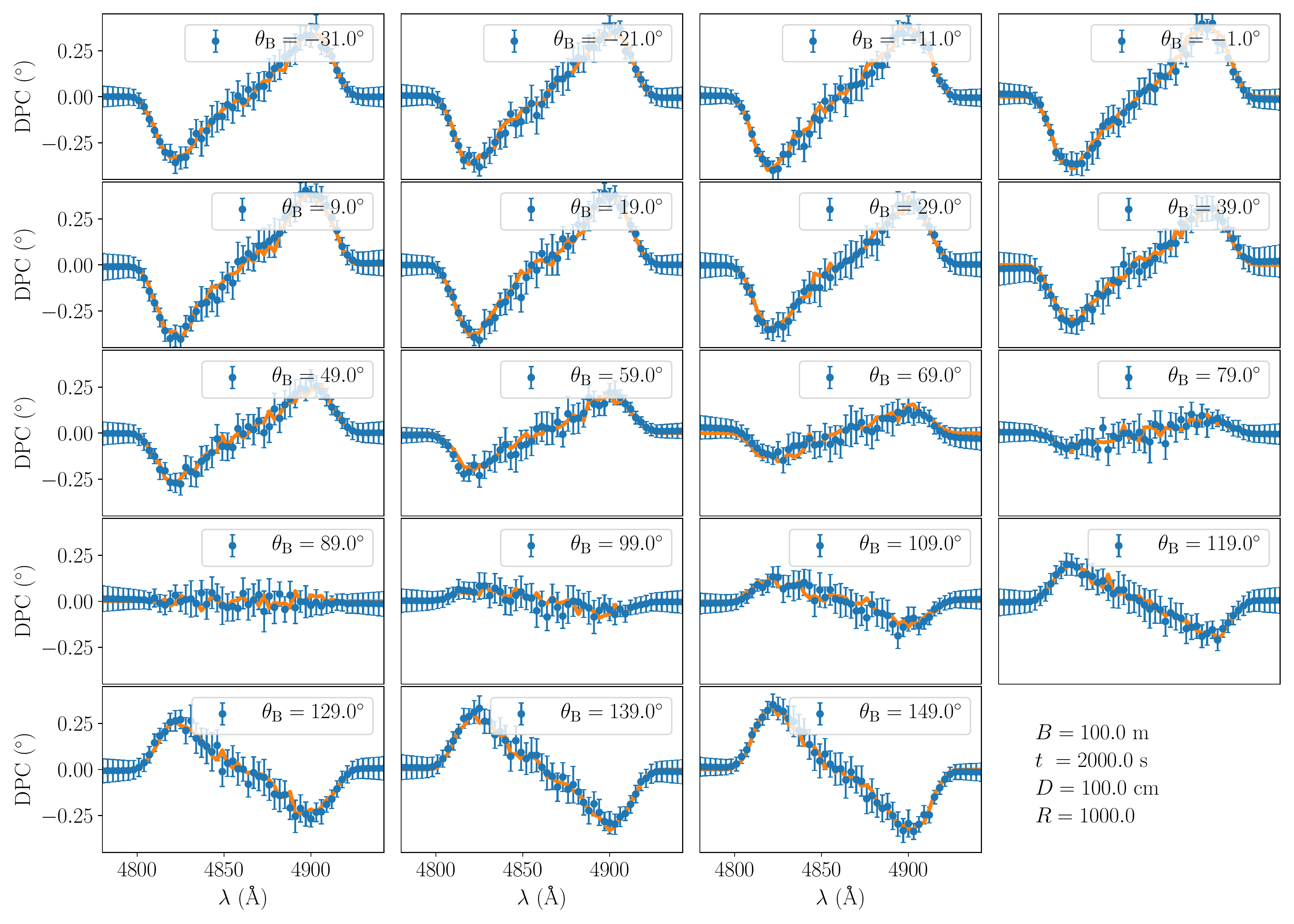}
   \end{minipage}
   \caption{An example of differential phase curves we generated differing only in baseline projected angle $\theta_{\rm B}$. The values of necessary interferometer parameters used are listed in the lower right part. Blue points with error bars are the differential phase. The thick solid lines are the best fitting using model parameters drawn from the probability distribution.  } 
   \label{fig:appendix1}
\end{figure}

\begin{figure}
\centering
\begin{minipage}[t]{1\textwidth}
  \centering
  \includegraphics[width=6.0cm]{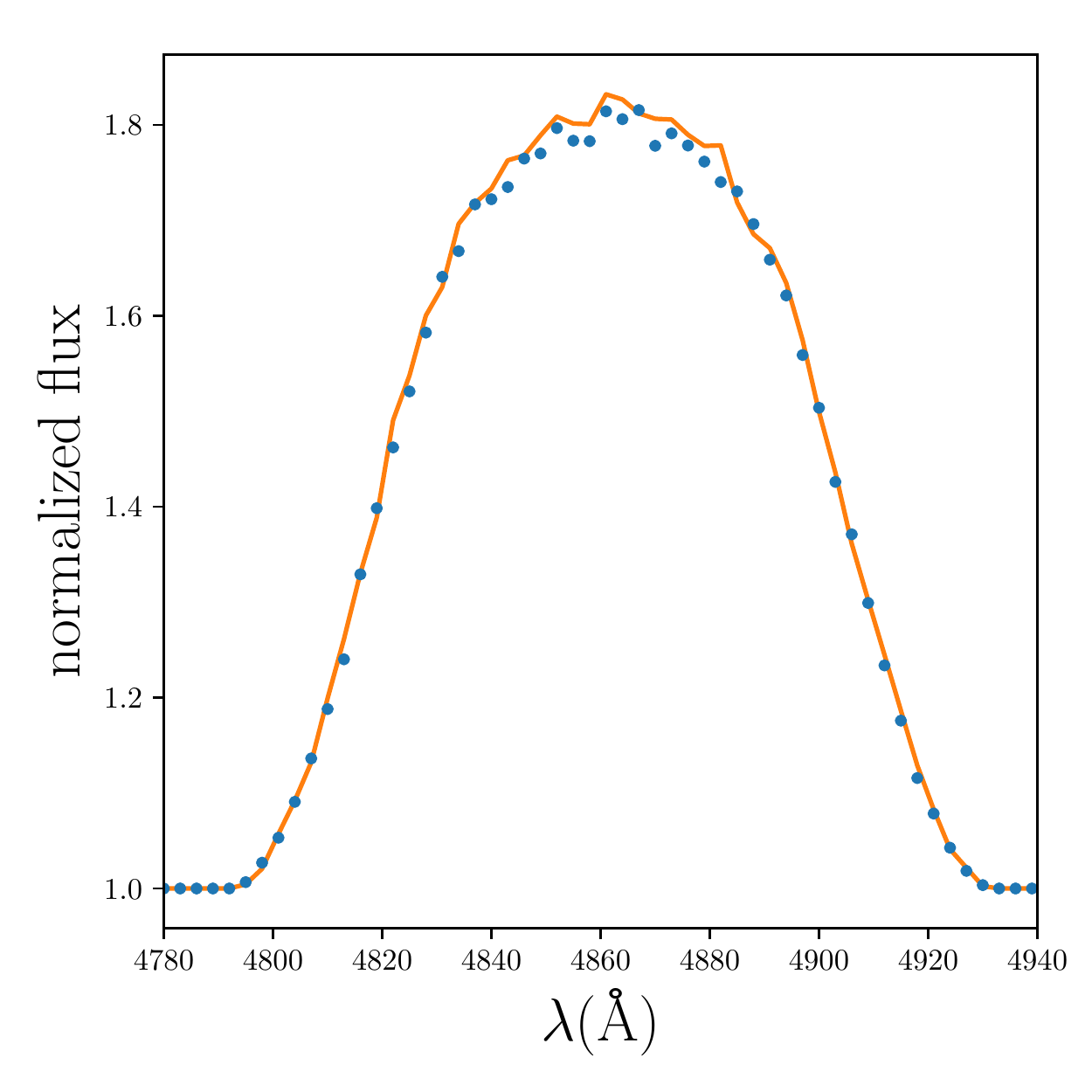}
  \caption{An example of H$\beta$ profile we generated (blue points). The thick solid line is the best fitting using model parameters drawn from the probability distribution.}
  \label{fig:appendix2}
\end{minipage}
\end{figure}

In this part, we use an example to illustrate the process of generating simulated data 
and the process of fitting. 
By adopting the parameterized BLR model described in Section \ref{sect:BLRmodel} and the values of parameters listed in Table \ref{table:blrpara},
we calculate the surface brightness distribution of BLR.
We used $2 \times 10^5$ (same as \citealt{Gravity2018}) clouds in the model.
Combined with the basic observational parameters and the values listed in Table \ref{table:interferometry},
we simulate the H$\beta$ emission line and DPCs.
The spectral resolution used is 1000, so there are 67 wavelength bins between  4750 $\rm{\AA}$ and 4950$\rm {\AA}$.
For each wavelength bin, we obtain the number of photons for each cloud that belongs to this bin.
Combined with the position of the cloud, we calculate the total flux and the
photocenter of the bin. 
and then calculate the total flux and the photocenter of the this bin. 
The mock H$\beta$ line profile (blue points) are shown in Figure \ref{fig:appendix2}.
We generate 19 mock DPCs by changing projected angle from 0$^{\circ}$ to 180 $^{\circ}$.
Blue points with errorbars in Figure \ref{fig:appendix1} are the mock data.

\begin{figure} 
   \centering
   \begin{minipage}{1.0\textwidth}
   \includegraphics[width=15.0cm, angle=0]{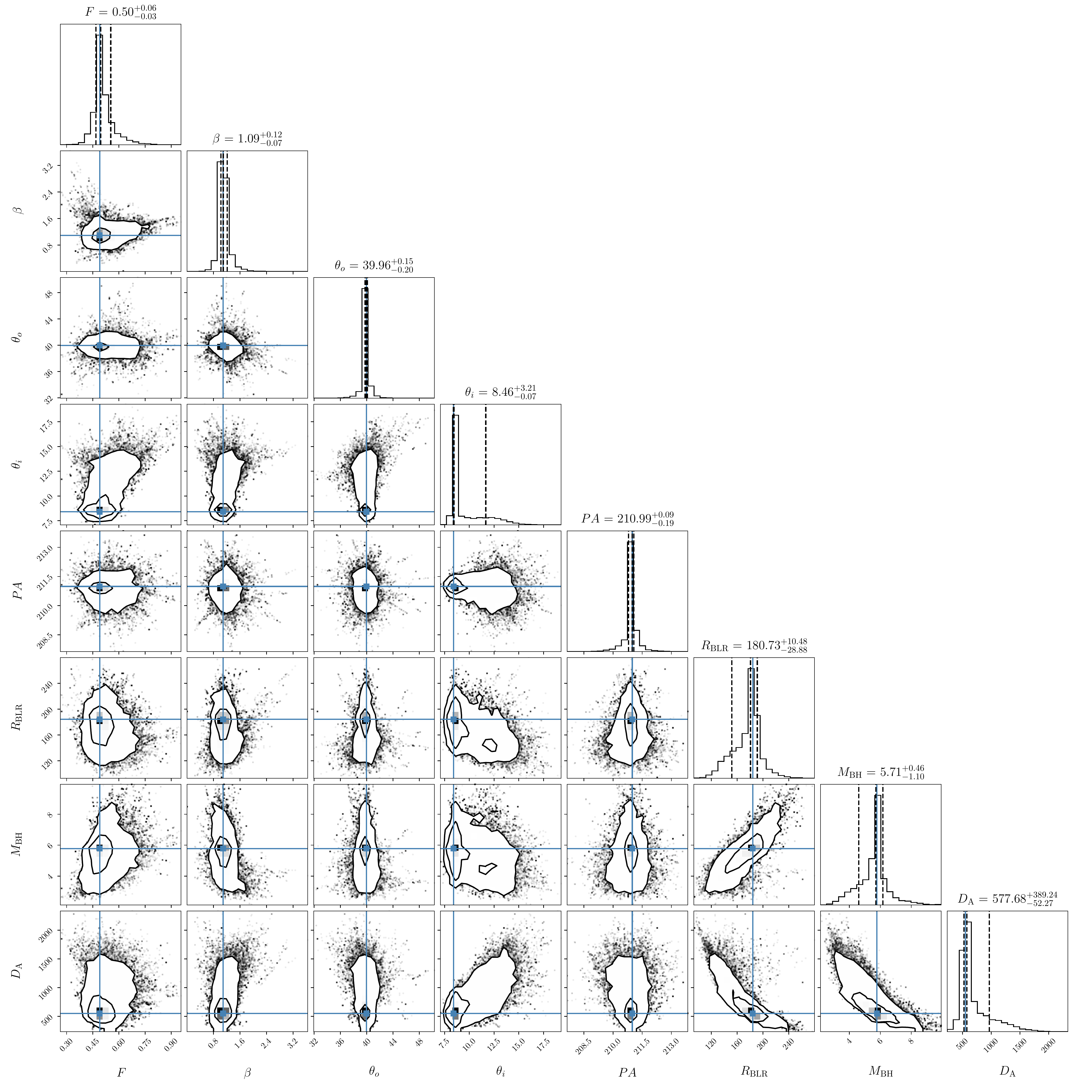}
   \end{minipage}   
   \caption{Probability density distributions of BLR parameters. The median values and error bars (1 $\sigma$ level) of the parameters are given on the top of panels. The blue lines represent input values. The contours are at 1 $\sigma$ and 2 $\sigma$ respectively. The dashed lines in the one-dimensional distributions are the 16$\%$, 50$\%$ and 84$\%$ quantiles.} 
   \label{fig:appendix3}
\end{figure}

We fit the BLR model to these mock DPCs. 
Bayesian statistics are used to measure confidence intervals of the model parameters. 
The priors are listed in Table \ref{table:blrpara}.
Markov Chain Monte Carlo code EMCEE is used to sample the posterior and obtain the posterior probability distribution of BLR model parameters. 
The results are shown in Figure \ref{fig:appendix3}. 
The median values of parameters are given on the tops of panels.

\acknowledgements
We are grateful to the referee for constructive suggestions to improve the manuscript.
We acknowledge the financial support of the National Natural Science Foundation of China (grants No. 12003077, 11703077, 12073068) and the Yunnan Province Foundation (202001AT070069).

\label{lastpage}

\end{document}